\documentclass[11pt,floatfix,a4paper,preprintnumbers,amssymb]{JHEP3}

\usepackage{amsmath}
\usepackage{graphicx}
\usepackage{graphics}
\usepackage{epsfig}
\usepackage{psfrag}
\usepackage{rotating}
\usepackage{dcolumn}
\usepackage{bm}

\newcommand{\beq}{\begin{eqnarray}} 
\newcommand{\eeq}{\end{eqnarray}}

\preprint{LPT-ORSAY-11-40 \\
CERN-PH-TH/2011-114}

\title{Forward-backward asymmetries of the bottom and top quarks in warped extra-dimensional models :  
LHC predictions from the LEP and Tevatron anomalies}

\author{Abdelhak Djouadi \\
Laboratoire de Physique Th\'eorique, CNRS and Universit\'e Paris-Sud 11, B\^at. 210
\\ F-91405 Orsay Cedex, France \\
Theory Unit, CERN, 1211 Gen\`eve 23, Switzerland \\
E-mail : \email{Abdelhak.Djouadi@cern.ch}}
\author{Gr\'egory Moreau \\
Laboratoire de Physique Th\'eorique, CNRS and Universit\'e Paris-Sud 11, B\^at. 210
\\ F-91405 Orsay Cedex, France \\
E-mail : \email{Gregory.Moreau@th.u-psud.fr}}
\author{Fran\c{c}ois Richard \\
Laboratoire de l'Acc\'el\'erateur Lin\'eaire, IN2P3/CNRS and Universit\'e Paris-Sud 11
\\ Centre Scientifique d'Orsay, B. P. 34, F-91898 Orsay Cedex, France \\
E-mail : \email{richard@lal.in2p3.fr}}

\abstract{Within the paradigm of warped extra dimensions, third generation quarks are expected to be the most sensitive
to effects beyond the Standard Model. The anomalies observed at the LEP
and Tevatron colliders in the forward-backward asymmetries of the bottom ($A_{\rm FB}^b$) and top ($A_{\rm FB}^t$) quarks 
can thus be seen as early signatures of warped extra-dimensional scenarios. 
We propose a realization of such a scenario, with a gauge custodial symmetry in the bulk, which
allows to address simultaneously the $A_{\rm FB}^b$ anomaly and the discrepancies observed recently on   
$A_{\rm FB}^t$ at high top quark rapidities and $t\bar t$ invariant masses. We then show that the various phenomenological constraints
arising from LEP, Tevatron and LHC can be satisfied within the considered model. The model   
predicts new features, induced by a Kaluza-Klein excitation of the gluon at a mass $\sim 1.5$--$2$~TeV,  
in top quark pair production at the $\sqrt{s}=7$~TeV LHC.}


\begin{document} 
\flushbottom

\subsection*{1. Introduction}
\label{intro}

There is increasing evidence that departures from the Standard Model (SM) are experimentally observed in the sector of third generation quarks.  
First, there is the longstanding anomaly of the forward-backward (FB)
asymmetry for $b$-quark jets, $A_{\rm FB}^b$, measured at 
LEP \cite{PDG} which differs by almost three standard deviations from the SM value 
at the $Z$ boson pole \cite{AFB-SM}. Then, the D0 \cite{D0} 
and CDF \cite{CDF} collaborations have reported
results on the FB asymmetry, $A_{\rm FB}^t$, in top quark
pairs produced at the Tevatron collider that are significantly higher than 
the SM expectation.
More recently, this excess has been confirmed by updated CDF data based on
an higher luminosity \cite{CDF11010034} which, interestingly, show that the excess  
in the $t\bar t$ rest frame appears mainly at high $t\bar t$ invariant masses ($M_{t\bar t}$) -- 
being at +3.4 standard deviations from the SM value above $M_{t\bar t}=450$~GeV -- as well as  
at high rapidities ($\Delta y$) -- being at +1.9 standard deviations from the SM for $\vert\Delta y\vert>1$.
This excess in $A_{\rm FB}^t$ has also been observed in the dilepton channel, i.e. when reconstructing the two top quarks 
from their leptonic decay (in contrast, the experimental data mentioned above result from investigations in the lepton+jets channel),
for $M_{t\bar t}>450$~GeV \cite{CDFdilepton}~: the non-unfolded result is at +2.6 standard deviations from the SM 
in the laboratory frame which cannot be directly compared to the lepton+jets channel but confirms a significant  
excess.

From a theoretical point of view, over the past decade, 
there have been intensive developments about an attractive alternative to supersymmetry~:  
the warped extra-dimension theory proposed by Randall and Sundrum (RS) \cite{RS}. The RS scenario   
possesses several deep motivations like the protection of the electroweak (EW) scale against radiative corrections.
Letting the SM fields propagate 
in the bulk allows to suppress higher dimensional operators 
and to generate the flavor structure (see e.g. Ref.~\cite{RSloc,RSmass,RSmassBIS})
if the heaviest SM fermions can be localized towards the so-called TeV-brane where the Higgs boson is confined. In this context, 
due to the large wave function overlap between the third generation fermions and the Kaluza-Klein (KK) excitations of gauge bosons 
(near the TeV-brane), the third generations fermions are expected to be the most
sensitive to new physics effects.  
For example, the significant bottom couplings to KK $Z$ type bosons $Z^{(n)}$ can lead to non-negligible corrections to the $Zb\bar b$ vertex,
via $Z$--$Z^{(n)}$ mixings. This feature allows to address the $A_{\rm FB}^b$ anomaly at LEP \cite{dmr}. 
Similarly, the effect of the KK gluon exchange at hadron colliders is expected to be large in the top quark sector. 
Indeed, the KK gluon exchange in the s-channel can soften \cite{dmrs} the discrepancy between the 
value of $A_{\rm FB}^t$ measured at the Tevatron \cite{CDF} and its SM value. 
Therefore, the anomalies observed in the third generation quark sector can be interpreted as early signatures of warped models.

In the present paper, we pursue our earlier efforts to explain these anomalies on $A_{\rm FB}^b$ \cite{dmr} and $A_{\rm FB}^t$ 
\cite{dmrs}\footnote{See Ref.~\cite{Leandro} for similar approaches in the dual models with a composite Higgs boson.}.  
We propose a new version of the RS model where the 
whole asymmetry $A_{\rm FB}^t$ is in perfect agreement with the updated CDF data \cite{CDF11010034} 
and the two new excesses of $A_{\rm FB}^t$ integrated above $M_{t\bar t}=450$~GeV and $\vert\Delta y\vert=1$ \cite{CDF11010034} 
are addressed\footnote{We note that effective approaches to these asymmetry excesses have been performed in Ref.~\cite{GrandeGrojean}.}.  
The main new ingredient is the choice of fermion localizations which allows for a parity violation in the first generation quark couplings 
and a low KK gluon mass, $M_{KK} \sim 1.5$~TeV, which leads to a significant $A_{\rm FB}^t$ enhancement. 
We have checked that the induced corrections leave the total and differential cross sections for $t\bar t$ production, measured
at the Tevatron and now at the Large Hadron Collider (LHC), as well as the EW precision observables in good agreement with the data.  
This scenario predicts an excess in the $M_{t\bar t}$ invariant mass distribution at the $\sqrt{s}=7$~TeV LHC 
that should be observed with a few fb$^{-1}$ data.

\subsection*{2. The theoretical model}
\label{formalism}

To protect the EW observables while allowing for not too heavy KK gauge bosons, $M_{KK} \sim 1.5$--$2$~TeV
[detailed discussion given in Section~4], we consider the bulk gauge custodial symmetry, ${\rm SU(2)_L \times}$ 
${\rm SU(2)_R \times \rm U(1)_X}$ \cite{ADMS},
which leads in particular to the presence of an extra $Z'$ boson at low energy.
The chiral quarks are promoted to the following universal representations under this symmetry group, 
looking e.g. at the third generation (see later for another possibility), one has, 
\begin{equation}
q_{1L}\!\in\! 
\left ( \begin{array}{ccc}
t_{1L} & b'_L & q'_{-4/3L} \\ b_{1L} & q''_{-4/3L} & q'_{-7/3L} 
\end{array} \right )_{-5/6} \  
q_{2L}\!\in\!  
\left ( \begin{array}{cc}
q'_{5/3L} & t_{2L} \\ 
t'_{L} & b_{2L} 
\end{array} \right )_{2/3} \  
b_R\!\in\!  
( b_R \ q'_{-4/3R} )_{-5/6} \ \ \
t_R\!\in\!  
( t_R )_{2/3}
\label{Represent}
\end{equation}
the subscript $-4/3$ of the exotic colored fermion $q'_{-4/3R}$ called custodian, for example, indicating its electric charge.
The $q_{1L}$ and $q_{2L}$ multiplets mix together on the Planck-brane resulting in the SM doublet $Q_L$
mainly composed here by the $q_{2L}$ component \cite{O3}; the 
universal mixing angle between the doublets is taken at $\sin^2\theta_{12}\simeq 2/3$.  

Solving the $A_{\rm FB}^b$ anomaly with a minimal $b_R$ multiplet imposes the embedding of $b_R$ in a $({\bf 1},{\bf 2})$ representation 
\cite{dmr}. From the point of view of $A_{\rm FB}^t$ optimization, in order to be able to greatly increase the $t_R$ coupling to the KK gluon,
we choose $t_R$ as a pure singlet so that there exist no light top partner custodians in the model and in turn no tight constraints from the 
direct searches of exotic quarks \cite{CustoBound}.
Then, the gauge invariance of the Yukawa couplings imposes to have two different multiplets $q_{1L}$ and $q_{2L}$ with the minimal choice
given in eq.~(\ref{Represent}).

The parameters $c_f$ fixing the 5-dimensional masses for each fermion $f$, $\pm c_f k$, $1/k$ being the $AdS$ curvature radius,  
control the fermion localizations in the bulk. We take for the three quark generations~:  
\begin{eqnarray}
& c_{u_L}=c_{d_L} \simeq 0.44, c_{u_R},c_{d_R} \simeq 0.80, 
c_{c_L}=c_{s_L} \simeq 0.62, c_{c_R} \simeq 0.62, c_{s_R} \simeq 0.49 
\nonumber \\
& c_{t_L}=c_{b_L} \simeq 0.51, c_{t_R}\simeq -1.30, c_{b_R}\simeq 0.53
\label{cset}
\end{eqnarray}
This complete set of parameters has been chosen to pass the various phenomenological
constraints and to address the bottom and top quark anomalies in the FB asymmetries, as will be discussed in details throughout the paper. These $c$
values also generate the correct order of magnitude for the lepton and quark masses through wave function overlaps with the
localized Higgs boson \cite{RSmass} (5-dimensional Yukawa coupling constants being of order $k^{-1}$)~: 
$m_t\simeq 175$~GeV,
$m_b\simeq 4$~GeV,
$m_c\simeq 0.3$~GeV,
$m_s\simeq 100$~MeV,
$m_u\simeq 3$~MeV and 
$m_d\simeq 5$~MeV.
The off-diagonal Yukawa couplings have not been specified and the $3\times 3$ flavor mixing treatment was not performed~: 
a precise fit of the masses together with the study of all the mixing angles is beyond the scope of this work.  
We simply assume sufficiently small off-diagonal couplings to the KK gluon to avoid dangerous 
flavor changing neutral current (FCNC)
effects\footnote{Let us simply mention that our set-up is close to the anarchical approach (5-dimensional Yukawa couplings $\sim k^{-1}$ and $c_{b,t}\lesssim 0.5$, $c_{\rm light}\gtrsim 0.5$) --
the only small differences being that $c_{u_L}=c_{d_L}\lesssim 0.5$ and $c_{s_R}\lesssim 0.5$ -- so that the fermion mixing 
angles should be generated as usually in the warped framework.
Regarding FCNC constraints, it is now admitted in the literature that an additional flavor structure might be
required to satisfy them, like introducing horizontal ${\rm U(1)}$ symmetries; see e.g. Ref.~\cite{FlavSym}.}.

\subsection*{3. The top quark at the Tevatron}
\label{Tevatron}

The FB asymmetry for the top quark at the Tevatron within the RS framework can be written in the initial $q \bar q$ center-of-mass frame 
in a form which allows to take into account the known SM contribution \cite{recastAFBt} at Next-to-Leading Order (NLO) in $\alpha_s$ 
\cite{Kuhn}~: 
\begin{eqnarray}
A_{\rm FB}^t = A_{\rm FB}^{\rm RS} \times R + A_{\rm FB}^{\rm SM} \times (1-R), \  \  \  
R = \frac{\sigma^{\rm total}_{\rm RS-LO}+\sigma^{\rm total}_{\rm inter.-LO}}
{\sigma^{\rm total}_{\rm SM-LO}+\sigma^{\rm total}_{\rm RS-LO}+\sigma^{\rm total}_{\rm inter.-LO}} ,
\nonumber 
\end{eqnarray}
\begin{eqnarray}
A_{\rm FB}^{\rm RS} = \frac{(\sigma^{\rm F}_{\rm RS-LO}+\sigma^{\rm F}_{\rm inter.-LO})-(\sigma^{\rm B}_{\rm RS-LO}+\sigma^{\rm B}_{\rm inter.-LO})}
{(\sigma^{\rm F}_{\rm RS-LO}+\sigma^{\rm F}_{\rm inter.-LO})+(\sigma^{\rm B}_{\rm RS-LO}+\sigma^{\rm B}_{\rm inter.-LO})}, \  \  \
A_{\rm FB}^{\rm SM} = \frac{\sigma^{\rm F}_{\rm SM-NLO}-\sigma^{\rm B}_{\rm SM-NLO}}
{\sigma^{\rm F}_{\rm SM-NLO}+\sigma^{\rm B}_{\rm SM-NLO}}, 
\nonumber
\end{eqnarray}
where 
$\sigma_{\rm SM-NLO}^{\rm F/B}$ is the cross section for the $t\bar t$ production in the SM at NLO integrated in the F/B hemisphere, 
$\sigma_{\rm RS-LO}^{\rm F/B}$ encodes the KK gluon exchange in the s-channel with $q \bar q$ initial state
and $\sigma_{\rm inter.-LO}^{\rm F/B}$ is the RS-SM interference part.
The contributions from the KK excitations of the neutral EW gauge bosons as well as the $Z'$ boson are negligible due
to the ratio of the EW over the QCD coupling and to the suppression of the $Z'\bar t_Rt_R$ vertex (with our $c$ values, $t_R$ 
is much more localized towards the KK $Z'$ profile than $t_L$) induced by the vanishing ${\rm SU(2)_R}$ isospin of $t_R$.
The $A_{\rm FB}^{\rm RS}$ and $R$ parts, computed at LO as indicated above, 
correspond to an approximated NLO calculation of the whole $A_{\rm FB}^t$ asymmetry since the K factors are approximately equal
in RS and in the SM \cite{RSnlo} so that they simplify in the ratios $A_{\rm FB}^{\rm RS}$ and $R$.

At the partonic level, the asymmetry for the main $q \bar q \to t \bar t$ process with $q$ denoting the isospin up or down type quark of the first generation,
which have the same couplings to the KK gluon [see eq.~(\ref{cset})], is at LO, 
\begin{eqnarray}
\hat A_{\rm FB}^{\rm LO} (\hat s) = a_qa_t \ \frac{4\pi\alpha_s^2(\mu_{\rm R})}{9} \frac{\beta^2_t \ |{\cal D}|^2\left[ (\hat s - M_{KK}^2) + 2 v_qv_t \ \hat s \right]}
{\hat \sigma^{\rm total}_{\rm SM-LO}(\hat s)+\hat \sigma^{\rm total}_{\rm RS+inter.-LO}(\hat s)}.
\label{partonicAFBt}
\end{eqnarray}
Here, $\hat s$ is the partonic squared energy, $\alpha_s$ is the QCD coupling 
at the renormalization scale $\mu_{\rm R}$, $\beta_t=\sqrt{1- 4m_t^ 2/\hat s}$
and the axial/vector couplings of the first KK gluon $g^{(1)}$ to the top quark and other light quarks $q$ are defined according to,
\begin{eqnarray}
a_q = \frac{1}{2}[Q(c_{q_R})-Q(c_{q_L})], \ 
v_t =  \frac{1}{2}[Q(c_{t_R})+Q(c_{t_L})], 
\label{Ax-Vec}
\end{eqnarray}
where $Q(c_f)$ quantifies the wave function overlap between two fermions $f$ and $g^{(1)}$, having $Q(+\infty) \simeq -0.2$.  
We have taken e.g. $c_{uL}<0.5$ keeping $c_{uR}>0.5$, which leads to $Q(c_{uL})\neq Q(c_{uR})$, namely $a_u\neq 0$.
The inverse propagator in eq.~(\ref{partonicAFBt}) reads  
\begin{eqnarray}
\frac{1}{{\cal D}}=\hat s -M_{KK}^2  + i \frac{\hat s}{M^2_{KK}} \sum_q \ \Gamma_{g^{(1)}\to q \bar q} M_{KK} \frac{\beta_q [v_q^2(3-\beta^2_q)/2+a_q^2\beta^2_q]}
{v_q^2+a_q^2}
\label{prop}
\end{eqnarray}
$\Gamma_{g^{(1)}\to q \bar q}$ denoting the $g^{(1)}$ partial width for decays into $ q \bar q$.  
Note that the energy dependence of the width, which is induced by 
radiative corrections to the $p\bar p \to t\bar t$ process,  
must be implemented for a significant total width
as is the case with our set of parameters~: $\Gamma_{g^{(1)}} \simeq 40\% M_{KK}$. 
This energy dependence leads to a shift in the pole position~:
\begin{eqnarray}
M_{KK} \to \ \ \sim \frac{M_{KK}}{(1+\Gamma_{g^{(1)}}^2/M_{KK}^2)^{1/4}} .
\label{shiftR}
\end{eqnarray}
In the kinematical region $\hat s \ll M^2_{KK}$ covered by the Tevatron, eq.~(\ref{partonicAFBt}) can be approximated by  
\begin{eqnarray}
\hat A_{\rm FB}^{\rm LO} (\hat s) \simeq -  \frac{a_qa_t}{M_{KK}^2} \ \frac{4\pi\alpha_s^2(\mu_{\rm R})}{9}  
\frac{\beta^2_t}{\hat \sigma^{\rm total}_{\rm SM-LO}(\hat s)},
\label{partonicAFBtsimpl}
\end{eqnarray}
taking into account the fact that the cross section is experimentally constrained to be close to its SM value.
We have chosen our parameters so that the axial quark couplings, $a_q\simeq -0.41$ and $a_t\simeq 3.41$, giving rise to a negative
$a_qa_t$ product, and the low mass $M_{KK}=1.5$~TeV maximize this asymmetry at LO which constitutes the 
RS contribution. In particular, the large $t_R$ coupling to the KK gluon combined with the small $t_L$ coupling 
lead to a maximized axial coupling $a_t$. At the next order in $\hat s / M^2_{KK}$, the $v_qv_t$ product is involved and turns out
to play a significant role~: it has to be minimized to increase the asymmetry.

\begin{figure}[!hc]
	\centering
	\vspace*{.5cm}
			\includegraphics[width=0.483\textwidth,height =5cm,angle=0]{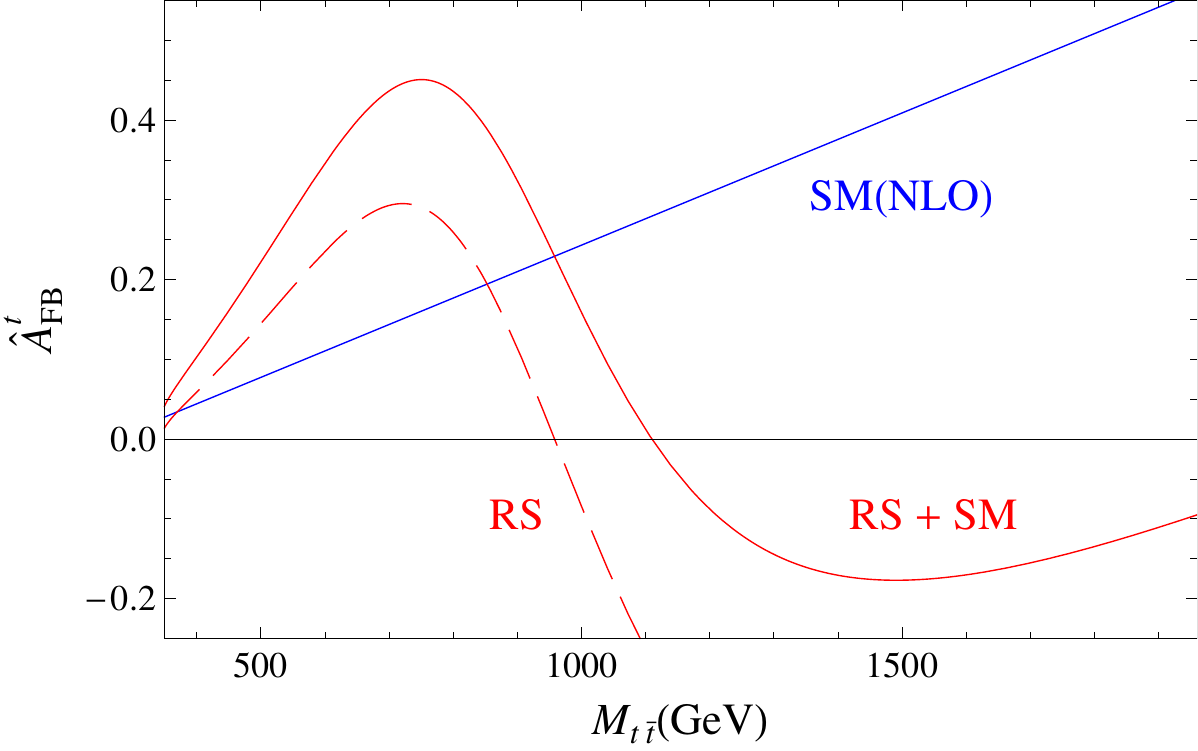}
			\includegraphics[width=0.47\textwidth,height =5.15cm,angle=0]{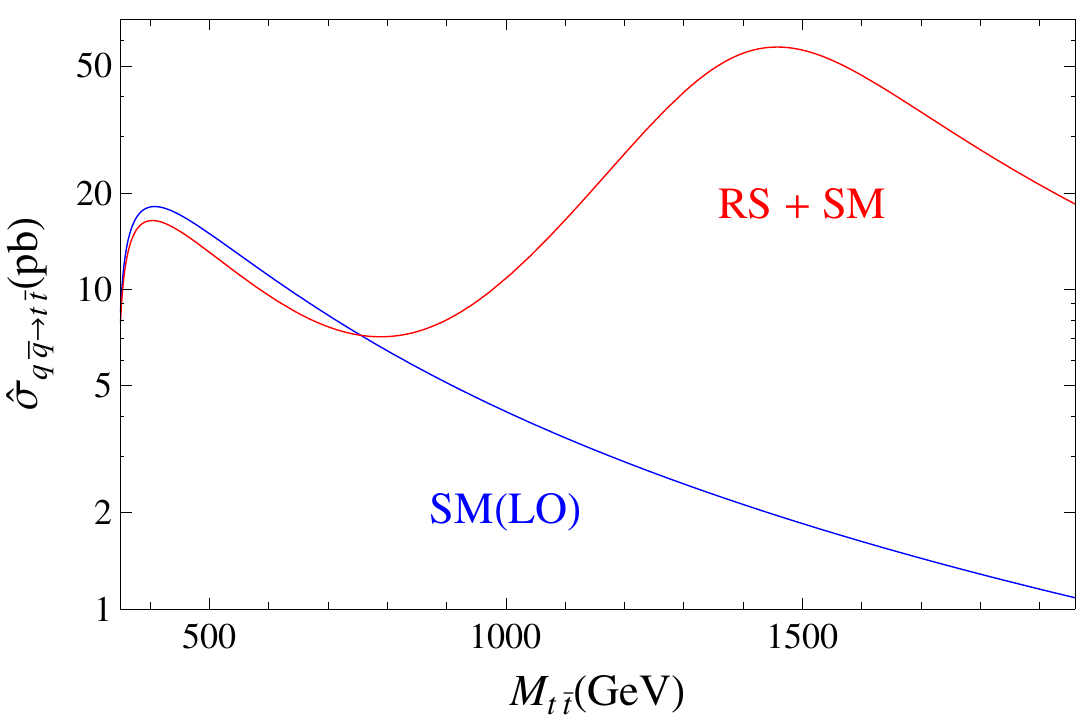}
\caption{\label{partonic} \small{Left~:  
The FB asymmetry for the partonic process
$q \bar q \to t \bar t$ as a function of the $t\bar t$ invariant mass (in GeV)
in the SM at NLO, $\hat A_{\rm FB}^{\rm SM-NLO}(\hat s)$ [blue line], at LO with the RS contribution, $\hat A_{\rm FB}^{\rm LO} (\hat s)$
[dashed red curve] and at NLO, $\hat A_{\rm FB}^{\rm NLO}(\hat s)$ 
(being the sum of the previous ones) [plain red curve]. We have taken $m_t=172.5$~GeV.
Right~: The partonic cross sections $\hat \sigma_{\rm SM-LO}$ (in pb) [blue curve] and 
$\hat \sigma_{\rm RS+SM+inter.-LO}$ [red curve] for $q \bar q \to t \bar t$.}}
\end{figure}

At NLO, one has the approximation, 
\begin{eqnarray}
\hat A_{\rm FB}^{\rm NLO} (\hat s) 
&=& \frac{(\hat \sigma^{\rm F}_{\rm SM-NLO}(\hat s)+\hat \sigma^{\rm F}_{\rm RS+inter.-LO}(\hat s))
-(\hat \sigma^{\rm B}_{\rm SM-NLO}(\hat s)+\hat \sigma^{\rm B}_{\rm RS+inter.-LO}(\hat s))}
{\hat \sigma^{\rm total}_{\rm SM-NLO}(\hat s)+\hat \sigma^{\rm total}_{\rm RS+inter.-LO}(\hat s)} 
\nonumber \\ 
& \simeq & \hat A_{\rm FB}^{\rm LO} (\hat s) + \hat A_{\rm FB}^{\rm SM-NLO} (\hat s).
\label{SumPart}
\end{eqnarray}
The partonic asymmetries $\hat A_{\rm FB}^{\rm NLO} (\hat s)$, $\hat A_{\rm FB}^{\rm LO} (\hat s)$ and $\hat A_{\rm FB}^{\rm SM-NLO} (\hat s)$ 
are drawn in the left-hand side of Fig.~\ref{partonic} 
as a function of $\sqrt{\hat s}=M_{t\bar t}$, $M_{t\bar t}$ being the $t\bar t$ invariant mass,  
for our set of $c$ parameters given in eq.~(\ref{cset}). We see on this figure that the RS contribution to the asymmetry in the mass
region relevant for the CDF data, $M_{t\bar t} \lesssim 800$~GeV, 
is positive and maximized which will lead to a better agreement of the asymmetries with data.

In the right-hand side of Fig.~\ref{partonic}, shown are the partonic cross sections~:
both in the SM, where the small component $gg\to t\bar t$ has been omitted for simplicity (but
its effect is included in all the other calculations) and
in the RS scenario where the $q\bar q \to g^{(1)} \to t\bar t$ channel is included. 
The resonance is visible at this partonic level, but
its effect will be suppressed when the cross section will be convoluted with
parton densities.

\begin{figure}[!hc]
	\centering
	\vspace*{.5cm}
			\includegraphics[width=0.481\textwidth,height =5cm,angle=0]{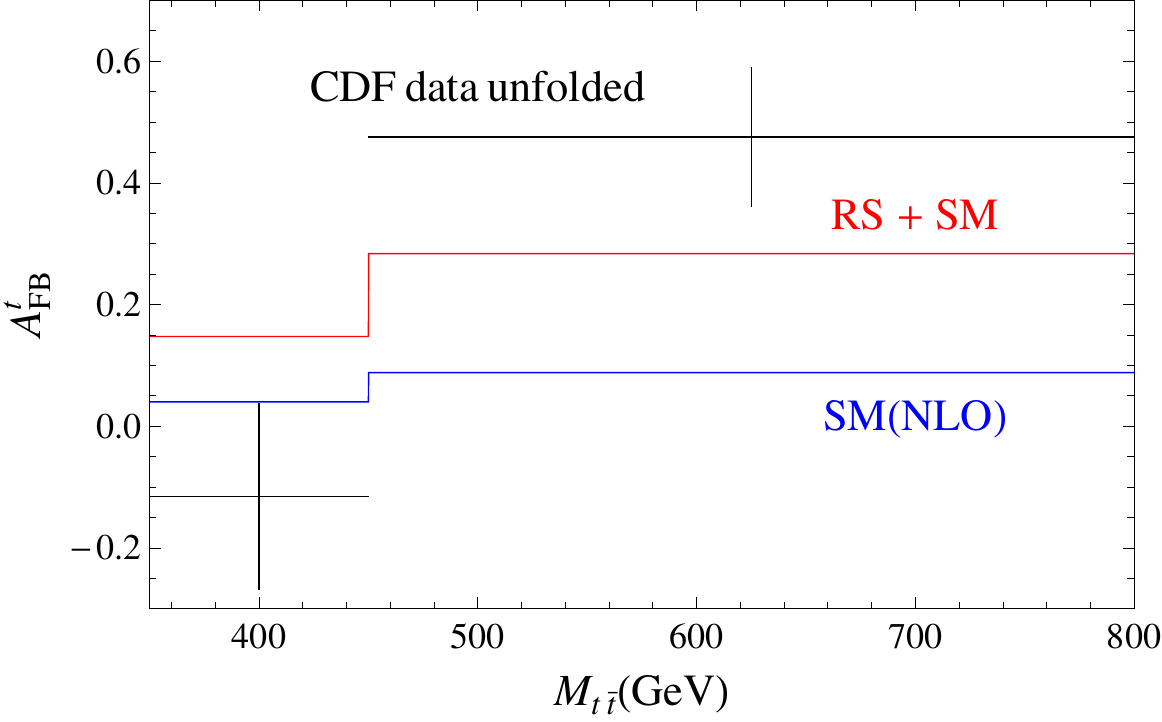}
			\includegraphics[width=0.49\textwidth,height =5cm,angle=0]{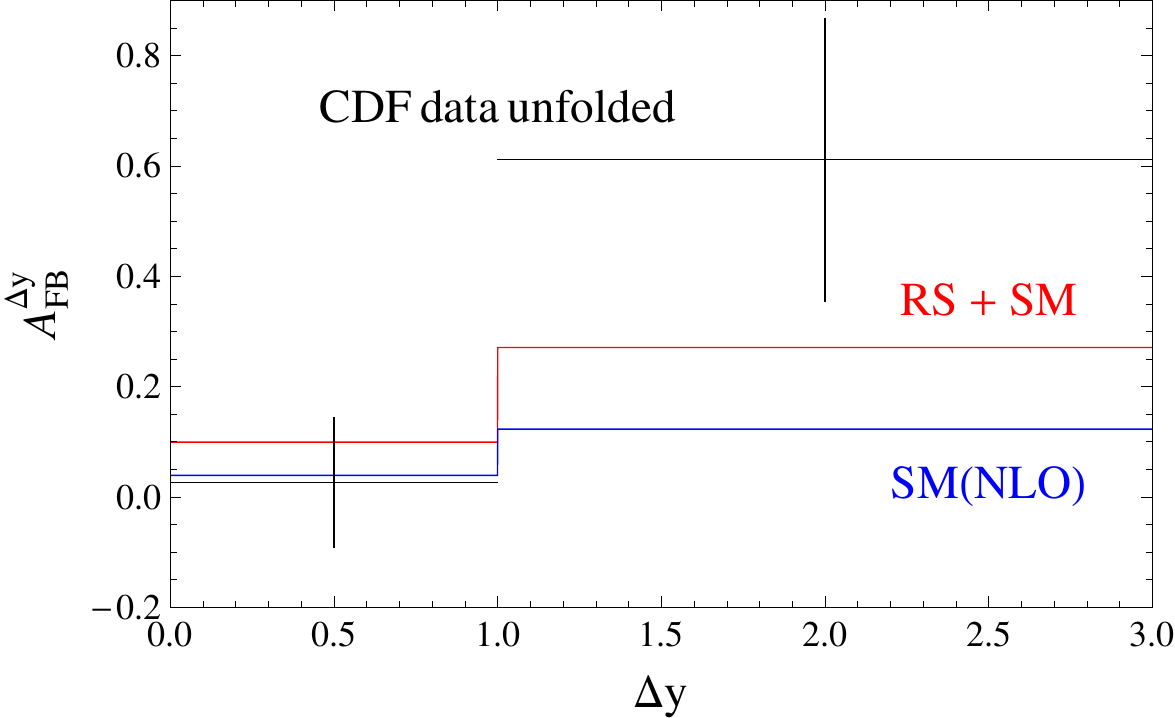}
\caption{\label{distrib} \small{Left~: The full top quark asymmetry  
integrated in the two energy ranges $[350,450]$ and $[450,900]$ 
of invariant mass $M_{t\bar t}$ (in GeV) computed within the RS extension 
of the SM, $A_{\rm FB}^t$, with $\mu_{\rm F}=\mu_{\rm R}=m_t=172.5$~GeV [red lines] 
and compared to the SM prediction at NLO, $A_{\rm FB}^{\rm SM}$  
[blue lines] as well as to the unfolded CDF data for $m_t=172.5$~GeV
\cite{CDF11010034} [black crosses for experimental errors]. 
In the first energy bin, $A_{\rm FB}^{\rm SM}$ is at $1.0\sigma$ from data
whereas $A_{\rm FB}^t$ is at $1.7\sigma$; in the second energy bin, $A_{\rm FB}^{\rm SM}$ is at $-3.4\sigma$ from data
whereas $A_{\rm FB}^t$ is away by $-1.7\sigma$. 
Right~: The asymmetries $A_{\rm FB}^{\vert\Delta y\vert<1}$ and $A_{\rm FB}^{\vert\Delta y\vert>1}$  
computed in the RS extension [red lines]  
and compared to the SM prediction at NLO  
[blue lines] as well as to the unfolded CDF data 
\cite{CDF11010034} [black crosses]. 
In the highest bin, $A_{\rm FB}^{\vert\Delta y\vert>1}$ in the SM is at $-1.9\sigma$ from data
whereas $A_{\rm FB}^{\vert\Delta y\vert>1}$ in RS is away by $-1.3\sigma$.}}
\end{figure}

Let us now come to the asymmetry once the partonic cross sections are folded with the parton density functions (PDF)
which we take from MSTW-2008-NLO \cite{MSTW09053531}.
As expected with the considered theoretical parameters, the left-hand side of Fig.~\ref{distrib}  
shows that the RS contribution increases the whole FB asymmetry, leading 
to a much better agreement with the recent unfolded CDF results\footnote{See Ref.~\cite{ERS} for discussions 
on the non-unfolded CDF data within a specific warped framework.} than in the pure SM case. 
The increase with $\hat s$ of the RS contribution to the asymmetry in Fig.~\ref{partonic} is responsible for a larger enhancement of $A_{\rm FB}^t$
above $M_{t\bar t}=450$~GeV in Fig.~\ref{distrib} as needed for a better fit to data. Nevertheless, choosing RS parameters that would lead to a larger
asymmetry above $M_{t\bar t}=450$~GeV would inevitably lead in the same time to an $A_{\rm FB}^t$ value too
far from the data below $450$~GeV. Hence, in the high $M_{t\bar t}$ range,
the improvement of $A_{\rm FB}^t$ from $-3.4$ standard deviations within the SM down to  
$-1.7$ standard deviations within the present RS model is the best improvement that one can hope in our warped higher-dimensional scenario.
This is due to an intrinsic constraint coming from the tension between the asymmetry measurements above and below $M_{t\bar t}=450$~GeV.
This feature suggests that the remaining discrepancy of $\sim 1.7 \sigma$
could be attributed to either a statistical fluctuation or higher order
QCD corrections. Indeed, the NLO corrections to the KK gluon exchange
could play a role in the present framework and, for the SM part,
additional and non-negligible contributions might arise at NNLO.

The fit on the
asymmetry $A_{\rm FB}^t$ integrated over the whole $M_{t\bar t}$ range is also greatly improved in our RS scenario compared to the SM case,
as one sees by comparing the theoretical prediction of our RS extension with the measurement [for $\mu_{\rm R}=\mu_{\rm F}=m_t=172.5 \ \mbox{GeV}$]~:
\begin{eqnarray}
\mbox{{\bf Tevatron data} \cite{CDF11010034}} &:& \ 0.158\pm 0.075 \nonumber \\ 
\mbox{{\bf SM [NLO]} \cite{CDF11010034}} &:& \ 0.058\pm 0.009 \ (-1.33\sigma) \nonumber \\ 
\mbox{{\bf RS+SM}} &:& \ 0.189\pm 0.010  \ (+0.42\sigma) \nonumber 
\end{eqnarray}
where the standard deviations of the central theoretical values relatively to the experimental value are given in brackets.
We have checked that the present RS theoretical predictions on the asymmetry are very stable against scale variation as well as PDF and top quark
mass uncertainties, a mere consequence of the fact that it is defined as a cross section ratio.
Hence, the error given above on $A_{\rm FB}^t$ is mainly due to the SM uncertainty.

The FB asymmetries at low ($\vert\Delta y\vert<1$) and high ($\vert\Delta y\vert>1$) top rapidities, $y_t=\Delta y/2$, 
have been measured by the CDF collaboration \cite{CDF11010034} with a rapidity cut $\vert\Delta y\vert<3$.  
The right-hand side of Fig.~\ref{distrib}, in which are given these unfolded results, illustrates that the fit to data is improved in the RS realization
compared to the SM situation. The reason is that if high absolute rapidities are selected, then large $\cos \theta^*$ values are considered,
$\theta^*$ being the scattering angle, so that the asymmetry generated by the KK gluon exchange is maximized 
\cite{dmrs}\footnote{More generally, these excesses in $A_{\rm FB}^t$ can be due to s-channel exchanges of color octet vector bosons, interfering 
with SM top quark production, with masses in the vicinity of the TeV scale. For instance, effective axigluon-inspired scenarios 
can cure these anomalies \cite{RohiniI,HewettAxi} (note that the parameter space explored here is not the same as the one considered 
in Ref.~\cite{HewettAxi} and our resonance width is larger) 
if one makes sure that the axigluon exchanges do not affect drastically the well behaved $t\bar t$ (differential) cross 
sections.}.

The total cross section for top quark pair production at the
Tevatron in our RS scenario, calculated with the program of
Ref.~\cite{HATHOR} which includes the approximate NNLO 
corrections to the SM contribution, is found to be
$\sigma(p \bar p \to t\bar t)=6.62 \pm 1$~pb for $\mu_{\rm R}=
\mu_{\rm F}=m_t=172.5$~GeV when the MSTW PDF set is adopted. The
combined uncertainty is from the scale variation, PDF and the top quark
mass which have been estimated according to the procedure given
in Ref.~\cite{ttbarUncert}. 
Given the uncertainties, the cross section value is in a good agreement with the
value measured at the Tevatron, $7.50 \pm 0.48$~pb \cite{CDFsigma}, again
obtained for $m_t = 172.5$~GeV. This agreement is essentially
due to the large mass and total width of the KK gluon resonance induced
by the significant $g^{(1)}\bar t_R t_R$ coupling, which lead
to only a small departure from the SM prediction.

\begin{figure}[!hc]
	\centering
	\vspace*{.5cm}
			\includegraphics[width=0.6\textwidth,height =6.cm,angle=0]{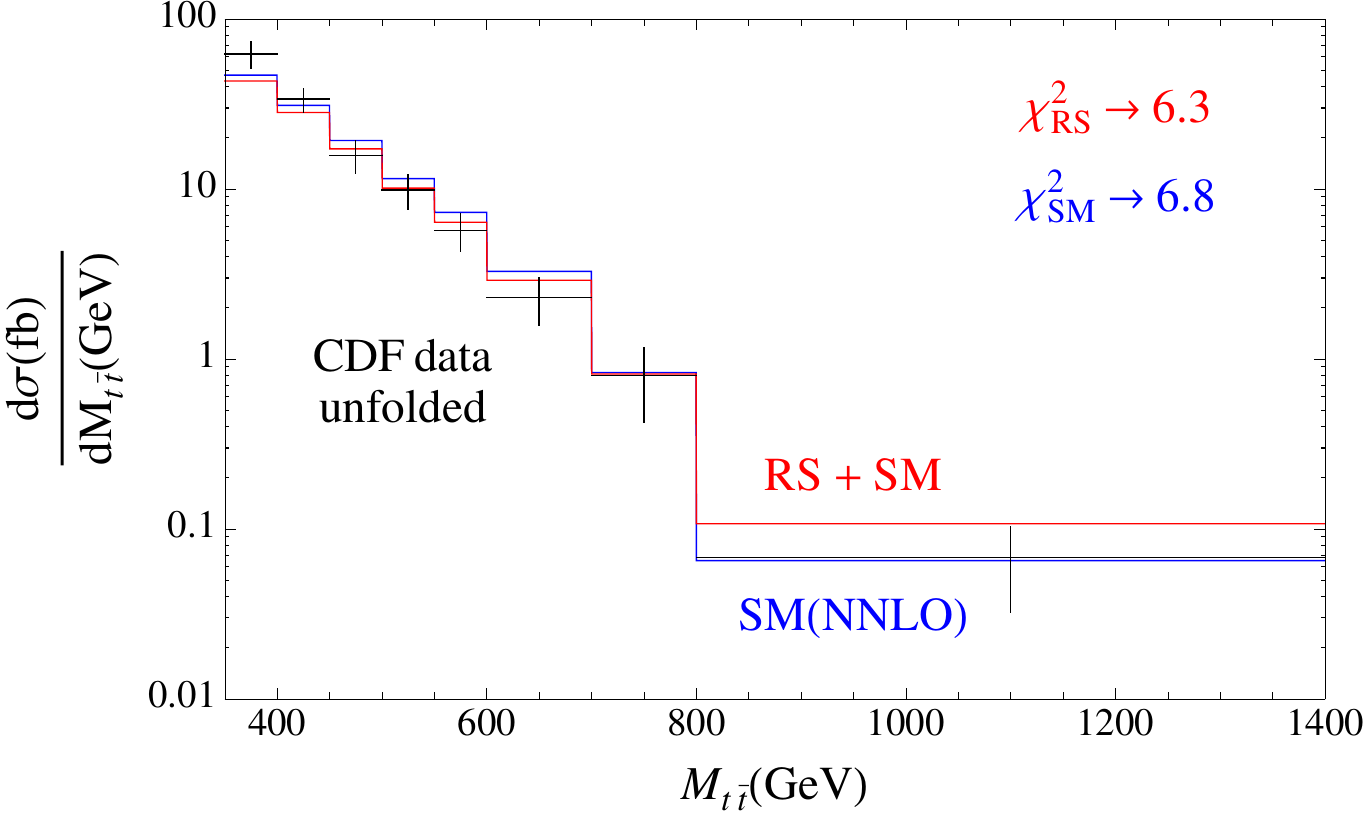}
\caption{\label{Xdistrib} \small{The differential cross section ${\rm d}\sigma_{\rm SM-NNLO}/{\rm d}M_{t\bar t}$ [in fb/GeV] 
at NNLO as a function of the $t\bar t$ invariant mass $M_{t\bar t}$ [in GeV] \cite{Neubert10035827} 
($\mu_{\rm F}=\mu_{\rm R}=m_t=175$~GeV) [blue curve] together with the distribution including the KK gluon exchange effect   
[red curve]. The unfolded CDF data of Ref.~\cite{CDF09032850} on these eight energy bins, for $m_t=175$~GeV, 
are also illustrated by the black crosses indicating the experimental error.
The differential cross section in the first bin is at $-1.4\sigma$ from the measurement within the SM whereas it lies at $-1.7\sigma$ in RS.
The resulting $\chi^2$ function values are indicated on the figure.}}
\end{figure}

An important final comment is on the $t\bar t$ invariant mass distribution
$$\frac{{\rm d}\sigma_{\rm SM-NNLO}}{{\rm d}M_{t\bar t}}(1+
\frac{{\rm d}\sigma_{\rm RS+inter.-LO}}{{\rm d}M_{t\bar t}}/\frac{{\rm d}\sigma_{\rm SM.-LO}}{{\rm d}M_{t\bar t}}),$$ 
with the SM part
normalized to the NNLO value, that is displayed in Fig.~\ref{Xdistrib} for our usual
set of parameters with $\mu_R=\mu_F=m_t=175$ GeV, using MSTW PDF.
Because of the destructive interference between the RS and SM contributions to the partonic cross section shown in Fig.~\ref{partonic}, 
the already existing excess relatively to the SM case in the first energy bin found by the 
CDF collaboration \cite{CDF09032850} becomes a little bit worse 
in our scenario as shows Fig.~\ref{Xdistrib}. 
However, in most of the subsequent energy bins, the KK exchange contribution improves the 
agreement with the data as is shown in Fig.~\ref{Xdistrib}. Hence, the global 
fit over the  considered eight energy bins is not degraded within our RS scenario~: 
the total $\chi^2$ function is $\chi^2_{\rm SM}=6.8$ 
in the SM while it decreases down to $\chi^2_{\rm RS}=6.3$ in RS.

\subsection*{4. LEP and other electroweak precision tests}
\label{LEPEWG}

The two precisely measured observables in the bottom sector are $A_{\rm FB}^b$ and 
$R_b= \Gamma( Z\to b\bar b)/\Gamma( Z\to {\rm hadrons})$ \cite{PDG}. 
With the choice of eq.~(\ref{Represent}) of representations and our $c$ values in eq.~(\ref{cset}), there is only one
$b'$ custodian per generation, contained in the $q_{1L}$ multiplet, whose smallest 5-dimensional parameter is
$c_{u_L}=c_{d_L} \simeq 0.44$; the corresponding minimal mass is
$\sim 1.5$~TeV and in turn this lightest $b'$ type 
custodian does not lead to very large $b-b'$ mixing effects on the $Z\bar b b$ vertex \cite{dmr}. The $Z$--$Z^{(n)}$
and $Z$--$Z'$ mixings induce important corrections to this vertex \cite{dmr}. The anomaly 
on $A_{\rm FB}^b$ can be cured, while keeping $R_b$ in good agreement with the LEP data, along the same lines as in the
warped frameworks of Ref.~\cite{dmr} (we consider a Higgs mass, $m_H$, close to its direct limit of  
$\sim 114$~GeV so that the lower bound
on $M_{KK}$ from EW precision tests can be minimized). 
For the set of 5-dimensional parameters considered here, the obtained deviations of 
$A_{\rm FB}^b$ and $R_b$ with respect to data are given in Table~\ref{MainTable},
for a $Z'$ coupling $g_{Z'}\simeq 2.6$, and both give rise to a good agreement. 

\begin{table}[!ht]
\vspace*{.5cm}
\begin{center}
\begin{tabular}{|c|c|c|c|c|c|c|c|c|c|c|}
\hline  
{\bf Obs.} & $A_{\rm FB}^b$ & $R_b$ & $A_{\rm FB}^c$ & $R_c$ & $A_{\rm FB}^s$ & $\Gamma_{\rm had}^{Z}$ & $\Gamma_{\rm tot}^{W}$ 
& $\langle Q_{\rm FB}\rangle$ & $C_{1u}\!+\!C_{1d}$ & $C_{1u}\!-\!C_{1d}$ \\ 
\hline  
{\bf SM}                & $2.7\sigma$ & $0.8\sigma$ & $0.9\sigma$ & $0.0\sigma$ & $0.6\sigma$ & $1.3\sigma$ & $0.2\sigma$ 
& $1.1\sigma$ & $0.2\sigma$ & $1.1\sigma$ \\
\hline
{\bf RS}                & $1.2\sigma$ & $1.2\sigma$ & $0.9\sigma $ & $0.5\sigma$ & $0.2\sigma$ & $1.0\sigma$ & $0.2\sigma$ 
& $0.1\sigma $ & $0.8\sigma$ & $0.1\sigma$ \\
\hline    
\end{tabular}
\end{center}
\caption{List of EW precision observables in the quark sector with their  
standard deviations [in absolute value] 
for the theoretical predictions with respect to experimental data in the SM, taken from Ref.~\cite{PDG}, and in our RS realization.  
The observables are defined as in Ref.~\cite{PDG}. 
In particular, the observable $\langle Q_{\rm FB}\rangle$ is the
asymmetry in the average charges over hemispheres of hadronic events measured at LEP;
the $C_{1u,d}$ encode the effective couplings between two leptons and two quarks tested in measurements of parity-violating 
electron scattering on nuclear targets (APV, PVES).}
\label{MainTable}
\end{table}

In order to generate a non-vanishing $a_q$ coupling, the first generation of quarks must be  
slightly closer to the TeV-brane than in geometrical set-ups considered 
usually\footnote{Even if other configurations, e.g. with $c_{u,c} < 0.5$ \cite{ERSorigin}, 
have been shown to pass the EW precision tests.}~: here, $c_{u_L}=c_{d_L}\lesssim 0.5$, 
whereas $c_{u_L}=c_{d_L} > 0.5$ is often considered to minimize the isospin up and down quark couplings to KK gauge bosons and in turn to reduce
the mixing-induced corrections to EW observables. Here, we suggest a new possibility~: these couplings can be slightly increased if there is a compensation 
between the corrections to the first and second generation quark couplings involved in some EW observables. Indeed, we have also chosen $c_{s_R}\lesssim 0.5$
so that the deviations of the $u_L,d_L$ couplings to the $Z$ boson compensate mainly the deviations of the $s_R$ 
coupling to protect the precisely-measured total hadronic decay width of the $Z$
boson, $\Gamma_{\rm had}^Z$, against too large deviations from its SM theoretical prediction. 
The $s_R$ quark being in a different multiplet from the $u_L,d_L$ one, it can possess 
a different ${\rm SU(2)_R}$ isospin leading to different $Z$--$Z'$ mixing-induced corrections to the $Z$ coupling.
In Table~\ref{MainTable}, we present the list
of EW precision observables in the quark sector, together with 
their standard deviations relatively to the data, in the SM and RS cases for $M_{KK} = 1.5$~TeV. We see that within RS, each observable is in a good
agreement with the corresponding measurement. 
Concerning the combined Tevatron (CDF+D0)
and HERA (ZEUS+H1) data on the $Zuu$/$Zdd$ couplings, the RS scenario studied here is also   
compatible with the present constraints on the vector and axial couplings \cite{11044590}.

We have focused on the EW fits in the quark sector as we consider new specific quark locations aimed at addressing the 
bottom and top FB asymmetries. We will not treat in detail the EW precision tests in the lepton and gauge boson sector but we describe here, two ways of 
obtaining acceptable fits within the present context. 
To have a low KK mass at $\sim 1.5$~TeV -- usually EW precision tests impose rather $M_{KK}\gtrsim 2$--$3$~TeV
for $c_{\rm lept.}>0.5$ \cite{RSreview} -- a first possibility is to achieve  
compensations between corrections to different lepton chiral couplings involved in some EW observables, 
as mentioned above for the bottom quark sector \cite{dmr}.
This can be done by taking some $c_{\rm lept.}$ values slightly smaller than $0.5$ so that the $Z'$ couplings are non-vanishing and can induce 
different $Z$ coupling corrections due to different ${\rm SU(2)_R}$ lepton isospins.
The oblique corrections implemented via the S parameter \cite{STU}, 
which usually include the direct corrections to lepton couplings after some redefinition valid for
$c_{\rm lept.}>0.5$ \cite{ADMS}, 
should now be treated separately and combined.  
The other possibility is already known~: if the leptons are decoupled from the KK gauge bosons by taking $c_{\rm lept.} \simeq 0.5$, the EW precision tests -- including
direct lepton vertex corrections as well as oblique corrections through the S parameter -- can be
satisfied for $M_{KK}\gtrsim 1.5$~TeV \cite{Decoupl,DecWells}.   
The other important oblique corrections enter via the T parameter which is protected at the tree level by the custodial symmetry.
At the loop level, the fermion exchange contributions depend on their multiplet embedding. If the gauge custodial symmetry is weakly broken in the bulk,
the T parameter is fixed at tree level by an effective gauge boson mass giving a sufficient freedom 
\cite{dmr}\footnote{Note that recently, a similar framework has been suggested \cite{Quiros} where a modifications of the $AdS$ metric near the infrared brane
allows a KK scale at 1 TeV without conflicting with EW precision tests.}.

\subsection*{5. LHC physics}
\label{LHCphys}

\subsubsection*{$t\bar t$ production cross section~:}

The $g^{(1)}gg$ coupling is zero at tree level due to the orthonormalization condition on the wave functions along the extra dimension
\cite{RSloc} combined with the flat profile of the gluon.
The loop induced coupling leads to small contributions to the $t\bar t$ production cross section at the LHC \cite{LHCbosonLOOP}.  
The contribution of the 
KK gluon exchange originates mainly from the $q \bar q$ initial state so that the rate $\sigma(p p \to t\bar t)$ in RS+SM is not significantly
different from that in the SM, whose major contribution at the LHC is coming from gluon-gluon fusion. 
As a consequence, the RS contribution to the $t\bar t$ rate is in good agreement with the recent LHC data.  
Indeed, the theoretical NNLO prediction in our RS model for the central $\sigma(p p \to t\bar t)$ value, 
based on Ref.~\cite{HATHOR} with $\mu_{\rm F}=\mu_{\rm R}=m_t=173$~GeV, is at $-0.81\sigma$ 
(the SM is at $-0.86\sigma$) from the ATLAS measurement, $180\pm 18.5$~pb \cite{ATLASwebpage}, and at $+0.36\sigma$ (SM at $+0.31\sigma$) 
from the CMS value, $158\pm 19$~pb \cite{CMSwebpage}; even without taking into account the QCD uncertainties, the agreement is 
thus satisfactory.

\subsubsection*{Search for dijet resonances~:}

The search for resonance bumps with the LHC detectors at $\sqrt{s}=7$~TeV and ${\cal L}=36$~pb$^{-1}$
allows to constrain the dijet production cross section \cite{dijetCMS,dijetDATA}.  
The ATLAS analysis does not assume the narrow width approximation, as previous ones, so that 
one can rescale the dijet production cross section via axigluon exchange \cite{axigluonORIG}, considered in the exclusion plot of Fig.~3 in Ref.~\cite{dijetDATA}, 
to the cross section via KK gluon exchange. One then obtains for our parameter set that the
dijet production cross section is in a region
clearly not excluded by the ATLAS constraint, namely at $\sigma_{\rm dijet} \times {\rm acceptance} \simeq 0.023$~pb 
whereas the upper bound is around $1$~pb at our resonance mass $M_{KK}=1.5$~TeV.
There is indeed a factor $\sim 100$ difference between the KK gluon and axigluon cross sections due to much smaller KK gluon 
couplings to light quarks\footnote{The pattern induced by quark mass hierarchies, 
that the top couplings to the KK gluon are typically larger than the first quark generation ($q$) couplings, allows one to have quite large $g^{(1)}\bar t_Rt_R$ couplings 
maximizing $A_{\rm FB}^t$ while keeping small $g^{(1)}\bar qq$ couplings in order to pass the dijet constraints.
This is not the case for the axigluon which has universal couplings to quarks, at least in the motivated
original model of Ref.~\cite{axigluonORIG} with a broken ${\rm SU(3)_L \times SU(3)_R}$ chiral color group.}.

\subsubsection*{Angular dijet distribution~:}

At the LHC, valence quark scattering allows to explore large jet-jet invariant mass ($M_{jj}$) values \cite{dijetDATA,dijetANGULAR}. 
While this process is insensitive to an s-channel exchange of a heavy gluon, it can probe its u- and 
t-channel contributions. Quark-quark scattering is governed by gluon 
exchange in the t-channel with a propagator varying like $1/\hat t$ while the exchange of a 
KK gluon varies like $1/(\hat t-M_{KK}^2)$ i.e. with a flatter $\hat t$ dependence as
long as $-\hat t \leq M_{KK}^2$. 
By selecting values of $-\hat t$ close to $M_{KK}^2$, one can therefore maximize the relative contribution of a massive gluon to   
reach the best sensitivity. In practice, one selects a large jet-jet mass final 
state with the highest scattering angle (this occurs for $\cos\theta^\star=0$, where $\theta^\star$ is 
the quark-quark scattering angle in the center-of-mass frame) which leads to $\hat t=-\frac{1}{2}M_{jj}^2$. 
For $-\hat t\sim M_{KK}^2$, one needs to select $M_{jj}=\sqrt{2}M_{KK}\sim2$~TeV 
which is precisely the end of the mass domain explored by ATLAS and CMS. Using 
Ref.~\cite{axigluonORIG} which provides quark-quark scattering distributions for a heavy axigluon and extending them to the case of a KK gluon, 
we have checked that the LHC limits \cite{dijetDATA,dijetANGULAR} are compatible with our scenario. Indeed,  
we find that our KK gluon with a mass $M_{KK}=1.5$~TeV induces less than $10\%$ deviation from the SM contribution\footnote{In 
contrast, the axigluon scenarios of Ref.~\cite{OctetA}, used in Ref.~\cite{CDF11010034} to reproduce the $A_{\rm FB}^t$ excess 
by assuming a large axial coupling to light quarks, give a $\sim 70\%$ deviation at $\cos\theta^\star=0$, incompatible with the LHC data.}.

\subsubsection*{Predictions at the LHC from the new warped model~:}
\label{LHCpred}

\begin{figure}[!hc]
	\centering
	\vspace*{.5cm}
			\includegraphics[width=0.488\textwidth,height =5cm,angle=0]{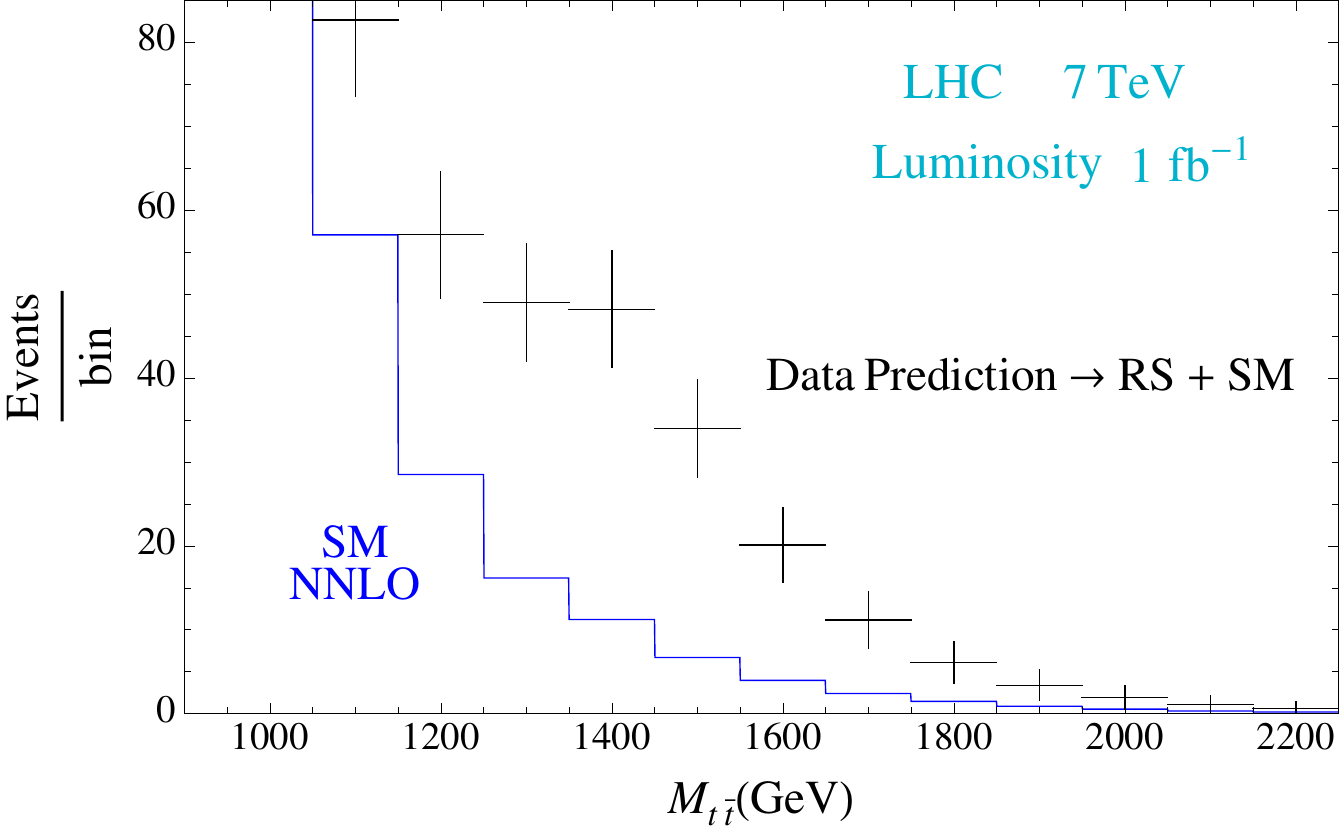}
			\includegraphics[width=0.488\textwidth,height =5cm,angle=0]{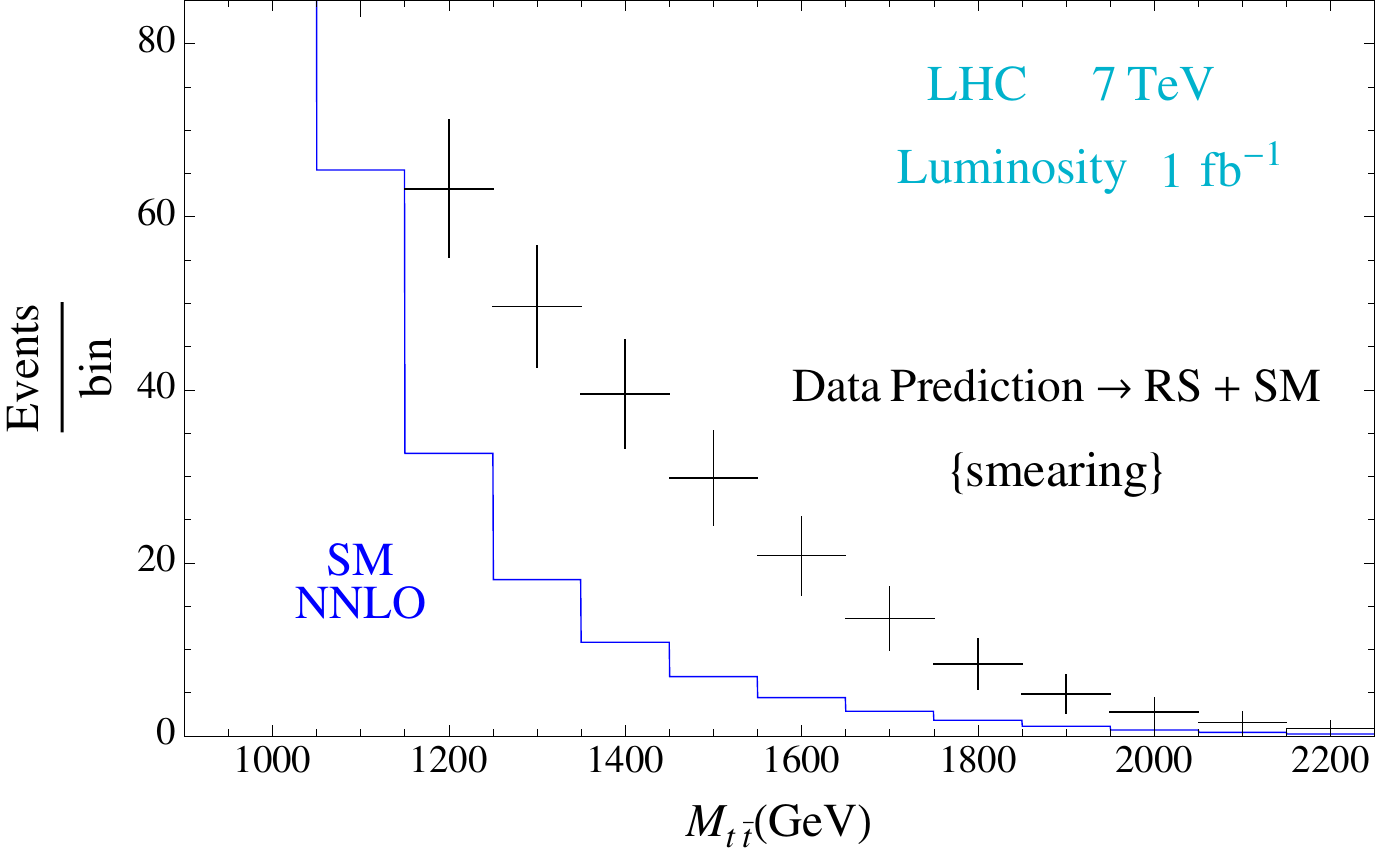}
\caption{\label{XdistribLHC} \small{The distributions of the invariant mass $M_{t\bar t}$ (in GeV) at the LHC 
assuming $100$~GeV bins with a luminosity of ${\cal L}=1$~fb$^{-1}$.
The SM at approximate NNLO ($\mu_{\rm F}=\mu_{\rm R}=m_t=173$~GeV) [blue histogram] \cite{HATHOR,Neubert10035827}  
is shown together with the RS contribution [in black~: the crosses indicate the statistical error]. 
The effect of smearing is only implemented in the right-hand side figure.}}
\end{figure}

The present RS realization, which resolves the $A_{\rm FB}^{b}$ and $A_{\rm FB}^{t}$ anomalies, would lead to striking effects in the $M_{t\bar t}$ distribution
that can be observed at the $\sqrt{s}=7$~TeV LHC with the expected luminosity of ${\cal L} \simeq 1$~fb$^{-1}$. 
These are an excess of events due to the large $g^{(1)}\bar t_Rt_R$ coupling and
a ``peak'' effect due to the KK gluon resonance, as shown in the left-hand side of Fig.~\ref{XdistribLHC}. 
Displayed are the $M_{t\bar t}$ distributions in the SM and the RS scenario with the bands indicating the statistical error. 
We have assumed a $t\bar t$ reconstruction efficiency of 
$\epsilon\approx 10\%$ as it lies between $5\%$ and $20\%$ depending e.g. on the top quark tagging \cite{ATLASeff}. 
The resonance bump observed in the left-hand side of Fig.~\ref{XdistribLHC} is not exactly peaked at $M_{t\bar t} = M_{KK}=1.5$~TeV, which is partly due
to the shift of the resonance induced by the energy dependence of the propagator, see eq.~(\ref{prop}) and eq.~(\ref{shiftR}); this shift is visible in the
right-hand side of Fig.~\ref{partonic}. The large total width $\Gamma_{g^{(1)}}$,   
combined with the sharp increase of the SM rate for decreasing $M_{t\bar t}$  
(where the parton densities are peaked), also tends to shift the resonance bump at low energies.

In the right-hand side of Fig.~\ref{XdistribLHC}, we have taken into account the fact that there is a finite experimental resolution in the 
measurement of the invariant mass $M_{t\bar t}$. For this purpose, we have convoluted the $M_{t\bar t}$ distributions with a gaussian function 
whose width is the resolution taken from Ref.~\cite{ATLASeff,LHCttlow}~: it is typically $\sim 150$~GeV at $M_{t\bar t}=1.5$~TeV. 
This results in a smearing effect of the distribution visible in the figure.

Hence, future LHC data should show a clear excess of events with respect to the SM; 
because of the smearing and the large KK gluon width, this resonance effect is not predicted to be a sharp peak in the 
$M_{t\bar t}$ distribution but the realistic shape, shown in the right-hand side of Fig.~\ref{XdistribLHC}, still clearly 
differs from the SM behavior.

We note that by rescaling the KK gluon distributions in Fig.~\ref{XdistribLHC} to the present LHC luminosity of ${\cal L} \simeq 36$~pb$^{-1}$,  
we obtain tiny numbers of events which are compatible with the available data \cite{LHCttlow}~: few or no events in the region
above $M_{t\bar t} \sim 1.1$~TeV where there is typically no background subtraction needed. 
The value of the $t\bar t$ production rate integrated above $M_{t\bar t}=1$~TeV 
in our scenario, normalized to the SM one, is $\sim 1.25$ as one deduces from the results obtained in the fifth reference of \cite{GrandeGrojean}
(Fig.~5 there) for $\Gamma = 0.1 M$ and couplings typically addressing the $A_{\rm FB}^t$ anomaly.

More challenging types of signals at LHC that could be manifestations of the present RS model are  
KK gluon effects in the dijet production, a second KK gluon excitation resonance -- predicted around $3.5$~TeV -- in the top pair production channel 
and anomalies in FB or charge asymmetries potentially measurable with a $pp$ initial state \cite{AFBLHC}.

\subsection*{6. Conclusion}
\label{conclu}

We have presented a RS scenario which allows for a common explanation of the anomalies observed at LEP and the Tevatron on the 
heavy quark forward-backward asymmetries, while satisfying the various tight constraints from collider data, including LHC. 
Our main conclusion is that such a scenario can be tested at the LHC with the sample of pair produced top quarks that will be collected this year. 
This data sample should
show a clear excess of events for $t \bar t$ invariant masses around $1.5$~TeV, induced by a KK gluon.

Let us stress that while several alternative models can also explain the excess in $A_{\rm FB}^t$, like in the case of axigluons for instance, 
they generally do not account for the $A_{\rm FB}^b$ anomaly. A possible way to
discriminate at the LHC the scenarios with the present KK gluon or an axi\-gluon is to search for KK electroweak 
gauge bosons -- e.g. via their decay into longitudinal $Z$ and Higgs bosons \cite{LHCboson} -- heavy quarks 
(custodians\footnote{Given our choice of a singlet representation for $t_R$ and the value of $c_{t_L}$ imposed by
phenomenological constraints, one predicts that all custodians should be heavier
than $\sim 1.5$~TeV. This scale is much above the experimental sensitivity of present
heavy quark searches which is about $400$~GeV. Morever, these searches assume a fourth
generation pattern for heavy quark decays \cite{CustoBound} while, e.g., a $b'$ custodian could decay predominantly into
$bZ$ and $bH$. Therefore, the actual mass limit for such particles could fall down to $\sim 150$~GeV,
if FCNC constraints are also satisfied. One could therefore
have chosen to embed $t_R$ into a doublet containing a light $b'_R$ custodian (taking the model of Ref.~\cite{dmrs}); the mass limit quoted
just above would have then imposed $c_{t_R}\gtrsim -0.5$ which translates into RS contributions to $A_{\rm FB}^t$ -- at high $M_{t\bar t}$ and $\Delta y$ -- 
lower than here, but still substantial. Thus, given the $A_{\rm FB}^t$ measurement accuracy, we cannot clearly rule out the presence
of a light $b'_R$ custodian. In such a case, the KK gluon width would significantly increase due to 
the opening of the channel $g^{(1)}\to \bar b'b'$, rendering more challenging the $g^{(1)}$ detection. On the other side, these colored custodians
would be abundantly produced at the LHC with a significant enhancement from the $g^{(1)}$ resonance contribution.}) \cite{ContServHou}
or a more challenging second KK gluon excitation (see also Ref.~\cite{RohiniII} for other discrimination me\-thods).   
\\
\\ 
\noindent \underline{Note added:} After our paper has been written, Ref.~\cite{LHCtt200} appeared where resonance searches in the
$M_{t\bar t}$ distribution from the $7$~TeV LHC at an integrated luminosity of ${\cal L} = 200$~pb$^{-1}$ lead to 
constraints on a KK gluon exchange. By rescaling the KK gluon production cross section and $t\bar t$ branching ratio 
in the Fig.~6 of Ref.~\cite{LHCtt200} to our KK gluon mass ($\sim 1.5$~TeV) and couplings, 
we obtain an acceptable $t\bar t$ production rate ($\sim 2.3$~pb) below the experimental upper limit 
($\sim 6$~pb)\footnote{Interestingly, in the Fig.~6 of  Ref.~\cite{LHCtt200},
the observed limit raises slowly up to $\sim 1.5\sigma$ from the expected limit
at $M_{KK}\simeq1.6$~TeV, and an event display has been chosen to be explicitly shown precisely for an high-mass candidate event at
$M_{t\bar t}\simeq1.6$~TeV.}. The KK gluon
width of the model \cite{ATLASeff} considered in Ref.~\cite{LHCtt200}, $\Gamma_{g^{(1)}} \simeq 15.3\% M_{KK}$, is smaller than here so that this  
constraint is even weaker in our case. 
\\
\\ 
\noindent \textbf{Acknowledgments} \ \  
We are grateful to Adam Falkowski and Fabienne Ledroit for fruitful discussions as well as to 
Valentin Ahrens and Julien Baglio for their precious help on QCD issues and to M.~Calvet for her work on the manuscript. 
This work is supported by the ERC Grant {\it Mass--TeV} as well as 
the ANR {\it CPV-LFV-LHC} under project \textbf{NT09-508531} and {\it TAPDMS} under project \textbf{09-JCJC-0146}.

%%%%%%%%%%%%%%%%%%%%%%%%%%

\end{document}